\documentclass{PoS}

\usepackage{epsfig}

\title{Dynamically induced scalar quark confinement:\\ A link between chiral
symmetry breaking and confinement}
\ShortTitle{Dynamically induced scalar quark confinement}

\author{\speaker{Reinhard Alkofer}\\
Institut f\"ur Physik,
Karl-Franzens-Universit\"at, 
Universit\"atsplatz 5,
A-8010 Graz, Austria\\
E-mail: \email{reinhard.alkofer@uni-graz.at}}
\author{Christian S. Fischer\\
        Institut f\"ur Physik, TU Darmstadt,
	Schlossgartenstr. 9,
	D-64289 Darmstadt, Germany}
\author{Felipe J. Llanes-Estrada\\
Universidad Complutense de
Madrid, Depto. F\'{\i}sica Te\'orica I. 28040 Madrid, Spain}
\author{Kai Schwenzer\\
Institut f\"ur Physik,
Karl-Franzens-Universit\"at, 
Universit\"atsplatz 5,
A-8010 Graz, Austria}

\abstract{Employing functional approaches
the infrared behaviour of Landau gauge QCD vertex functions is investigated. 
Results for the ghost-gluon, three-gluon and 
quark-gluon vertex functions are presented. 
As can be analytically shown a linear rising potential bet\-ween heavy quarks is
generated by infrared singularities in the dressed quark-gluon vertex. 
The selfconsistent mechanism that generates these singularities implies the 
existence of scalar Dirac amplitudes of the full vertex and the quark 
propagator. 
These amplitudes can only be present when chiral symmetry is broken, 
either explicitly or
dynamically. The corresponding relations thus constitute a novel mechanism that
directly links chiral symmetry breaking with confinement.}

\FullConference{The XXV International Symposium on Lattice Field Theory\\
		 July 30-4 August 2007\\
		 Regensburg, Germany}

\begin{document}

\section{Introduction}
 
Quark confinement and dynamical chiral symmetry breaking are the two most
prominent phenomena of infrared QCD. Recent Monte-Carlo lattice calculations
made clear that there is, at least for quarks in the fundamental representation,
a close and yet not fully understood relation between these two properties of
QCD. {\it E.g.\/} the spectral properties of the Dirac operator reflect both,
confinement and chiral symmetry breaking \cite{Gattringer:2006ci}. It is the
central aim of this talk to shed light onto this issue from the point of view of
QCD Green functions in the Landau gauge.

In this and related functional approaches dynamical chiral symmetry breaking
finds a direct explanation \cite{Roberts:1994dr}, the main challenge for such
non-perturbative methods is posed by the properties of the linearly rising
static quark-antiquark potential. There have been many quite different attempts
to relate the properties of this potential to properties of QCD, and thus 
explain quark confinement. In  ref.\ \cite{Alkofer:2006fu} some of these
pictures 
%where quark confinement is related to
%\begin{itemize}
%\addtolength{\itemsep}{-3mm}
%\item the condensation of chromomagnetic monopoles \cite{Mandelstam,Pisa},
%\item the percolation of center vortices \cite{Greensite},
%\item the AdS$_5$ / QCD correspondence \cite{Maldacena:1998im},
%\item the Gribov-Zwanziger scenario in Coulomb gauge \cite{Gribov,Dan1}, or
%\item the infrared behaviour of Landau gauge Greens functions
%\cite{Alkofer:2000wg,Fischer:2006ub,Alkofer:2006jf},
%\end{itemize}
have been briefly reviewed.
These explanations for confinement are seemingly
different but there are surprising relations between them which are not yet
understood. Given the current status one has to note that these theories are
definitely not mutually exclusive but simply reveal only different aspects of
the confinement phenomenon. And into one special facet, the above mentioned 
relation to broken chiral symmetry, there is novel insight from the Landau gauge
Greens functions approach.

\section{Infrared Yang-Mills theory in the Landau gauge\label{IRYM}}
\subsection{Infrared Exponents of Gluons and Ghosts}
\begin{figure}[th]
\centerline{\psfig{file=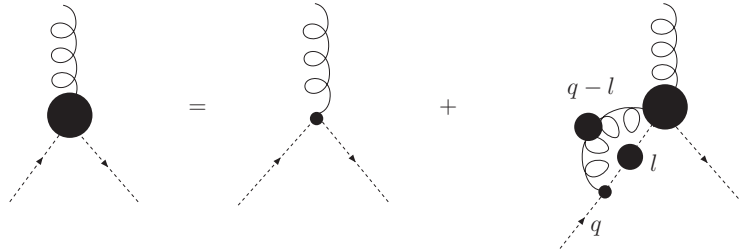,width=100mm}}
\vspace*{-8pt}
\caption{The Dyson-Schwinger equation for the ghost-gluon vertex.
\label{GhGlDSE}}
\end{figure}
Let us start by looking at the Dyson-Schwinger 
equation for the ghost-gluon vertex function as depicted in fig.\ \ref{GhGlDSE}.
In the Landau gauge the gluon propagator is transverse, and therefore one can
employ the relation
\begin{equation}
l_\mu D_{\mu \nu}(l-q) = q_\mu D_{\mu \nu}(l-q) \, ,
\end{equation}
to conclude that the ghost-gluon vertex stays
finite when the outgoing ghost momentum vanishes, {\it i.e.} when $q_\mu
\rightarrow 0$~\cite{Taylor:1971ff}. This argument is valid to all orders in
perturbation theory, a truely non-perturbative justification of the 
infrared finiteness of this vertex has been given in refs.\
\cite{Lerche:2002ep,Cucchieri:2004sq,Schleifenbaum:2004id}.

\begin{figure}[th]
\centerline{\psfig{file=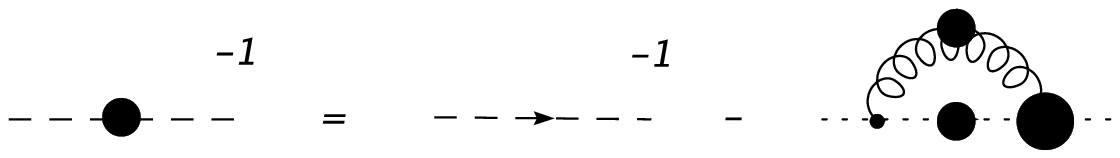,width=100mm}}
\vspace*{-8pt}
\caption{The Dyson-Schwinger equation for the ghost propagator.
\label{GhDSE}}
\end{figure}

Using this property of the ghost-gluon vertex
the Dyson-Schwinger equation for the ghost propagator,
see fig.\ \ref{GhDSE}, can be analysed.
The only unknowns in the deep infrared are the gluon and the ghost propagators:
\begin{eqnarray}
        D_{\mu\nu}(k) = \frac{Z(k^2)}{k^2} \, \left( \delta_{\mu\nu} -
        \frac{k_\mu k_\nu}{k^2} \right)  \; ,\quad
        D_G(k)  &=& - \frac{G(k^2)}{k^2}          \;.
\end{eqnarray}
In Landau gauge these (Euclidean) propagators are parametrized by
two invariant functions, $Z(k^2)$ and $G(k^2)$, respectively.
As solutions of renormalized equations,  these functions 
depend also on the renormalization  scale $\mu$. Furthermore, assuming that 
QCD Green functions can be expanded in asymptotic series, the integral in the
ghost Dyson--Schwinger equation can be split up in three pieces: an infrared
integral, an ultraviolet integral,  and an expression for the ghost wave
function renormalization. Hereby it is the resulting equation for the latter
quantity which allows one to extract definite information \cite{Watson:2001yv}
without using any truncation or ansatz.

One obtains that the infrared behaviour of the gluon and ghost propagators is
given by power laws, and that the exponents are uniquely related such that the
gluon exponent is -2 times the ghost exponent \cite{vonSmekal:1997is}. As we
will see later on this implies an infrared fixed point for the corresponding
running coupling. The signs of the exponents are such that the gluon propagator
is infrared suppressed as compared to the one for a free particle, the ghost
propagator is infrared enhanced.

\begin{figure}[th]
\centerline{\psfig{file=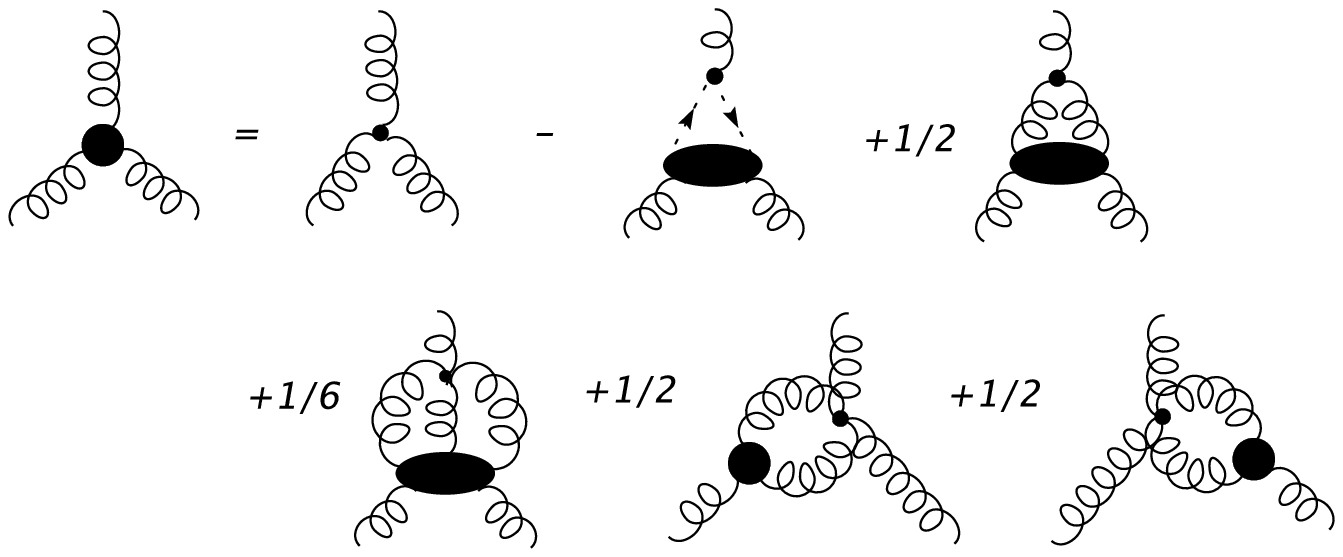,width=100mm}}
\vspace*{-8pt}
\caption{The Dyson-Schwinger equation for the 3-gluon vertex.
\label{3GDSE}}
\end{figure}
\begin{figure}[th]
\centerline{\psfig{file=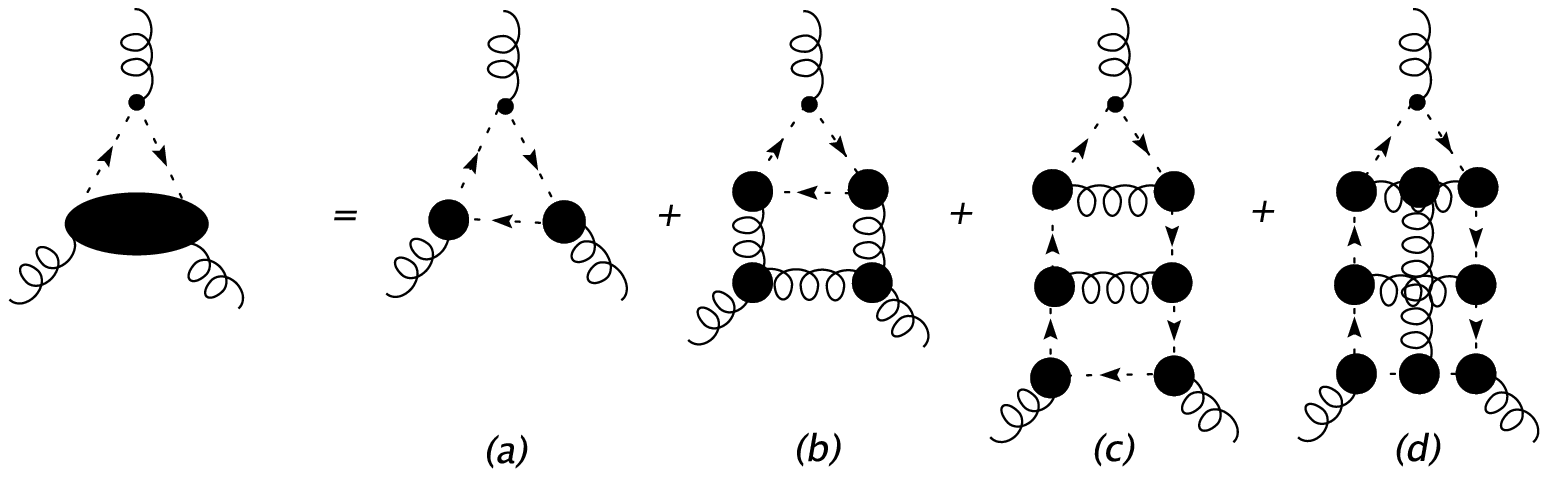,width=90mm}}
\vspace*{-8pt}
\caption{An example for the skeleton expansion of the
3-gluon vertex.
\label{3Gskel}}
\end{figure}
\begin{figure}[th]
\centerline{\psfig{file=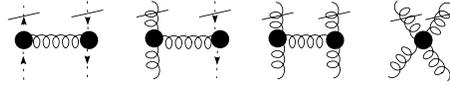,width=60mm}}
\vspace*{-8pt}
\caption{Insertions to reconstruct higher orders in the skeleton expansion.
\label{3GskelIn}}
\end{figure}

Given the infrared power laws, that the Yang-Mills propagators obey, one can infer 
the infrared behaviour of higher $n$-point functions. To this
end the  $n$-point Dyson-Schwinger equations have been studied in a
skeleton expansion, {\it i.e.\/} a loop expansion using dressed propagators and
vertices. Furthermore, an asymptotic expansion has been applied to all primitively
divergent Green functions \cite{Alkofer:2004it}.
As an example consider the Dyson-Schwinger equation for the 3-gluon vertex which
is diagrammatically represented in fig.\ \ref{3GDSE}.
Its skeleton expansion, see fig.\ \ref{3Gskel}, can be constructed
via the insertions given in fig.\ \ref{3GskelIn}.
These insertions have vanishing infrared anomalous dimensions which implies
that the resulting higher order terms feature the same infrared scaling.
Based on this
the following general infrared behaviour for one-particle irreducible
Green functions with $2n$ external ghost legs and $m$ external
gluon legs can be derived\cite{Alkofer:2004it,Huber:2007}:
\begin{equation}
\Gamma^{n,m}(p^2) \sim (p^2)^{(n-m)\kappa + (1-n)(d/2-2)}  \label{IRsolution}
\end{equation}
where $\kappa$ is one yet undetermined parameter, and $d$ is the
space-time dimension.
Exploiting
Dyson-Schwinger equations and Exact Renormalization Group Equations
one can show that this infrared solution is unique \cite{Fischer:2006vf}.

\subsection{Infrared fixed point of the Yang-Mills running coupling}

The infrared behaviour (\ref{IRsolution}) especially includes
\begin{equation}
{ G(p^2) \sim (p^2)^{-\kappa}}\;,
\quad { Z(p^2) \sim (p^2)^{2\kappa}}            
\quad {  \Gamma^{3g}(p^2) \sim (p^2)^{-3\kappa}} \;,
\quad { \Gamma^{4g}(p^2) \sim (p^2)^{-4\kappa}}
\end{equation}
and therefore the running couplings related to these
vertex functions possess an infrared fixed point:
\begin{eqnarray}
\displaystyle \alpha^{gh-gl}(p^2) &=&
\alpha_\mu \, { G^2(p^2)} \, { Z(p^2)}
\sim \frac{const_{gh-gl}}{N_c} ,
\quad
\displaystyle \alpha^{3g}(p^2) =
\alpha_\mu \, { [\Gamma^{3g}(p^2)]^2} \, { Z^3(p^2)}
\sim \frac{const_{3g}}{N_c} ,
\nonumber \\
\displaystyle \alpha^{4g}(p^2) &=&
\alpha_\mu \, {  \Gamma^{4g}(p^2)} \, { Z^2(p^2)}
\sim \frac{const_{4g}}{N_c}.
\end{eqnarray}
The infrared value of the coupling related to the ghost-gluon
vertex can be computed\cite{Lerche:2002ep,Fischer:2002hn}:
\begin{equation}
\alpha^{gh-gl}(0)=\frac{4 \pi}{6N_c}
\frac{\Gamma(3-2\kappa)\Gamma(3+\kappa)\Gamma(1+\kappa)}{\Gamma^2(2-\kappa)
\Gamma(2\kappa)}
\end{equation}
This yields $\alpha^{gh-gl}(0)=2.972$ for $N_c=3$ and $\kappa = (93 -
\sqrt{1201})/{98} \simeq 0.595353 $, which is the value obtained with a bare
ghost-gluon vertex.

\subsection{Positivity violation for the gluon propagator}

Positivity violation of the propagator of transverse gluons has
been for a long time a conjecture which has been supported recently,
see  {\it e.g.\/} \cite{Alkofer:2003jj,Bowman:2007du} and references
therein. The basic feature is hereby the infrared suppression of transverse 
gluons caused by the infrared enhancement of ghosts. 
As this behaviour clearly signals the confinement of tranverse gluons
\cite{vonSmekal:2000pz} it is certainly worth to have a closer look at the
underlying analytic structure of the gluon propagator.

Note that the infrared exponent $\kappa$ is an irrational number. Given the
infrared power laws this implies already that the
gluon propagator possesses a cut on the negative real $p^2$ axis. It is possible
to fit the solution for the gluon propagator quite accurately without
introducing further singularities in the complex $p^2$
plane \cite{Alkofer:2003jj}:
\begin{equation}
Z_{\rm fit}(p^2) = w \left(\frac{p^2}{\Lambda^2_{\tt QCD}+p^2}\right)^{2 \kappa}
 \left( \alpha_{\rm fit}(p^2) \right)^{-\gamma} .
 \label{fitII}
\end{equation}
$w$ is a normalization parameter, and
$\gamma = (-13 N_c + 4 N_f)/(22 N_c - 4 N_f)$
is the one-loop value for
the anomalous dimension of the gluon propagator.
The running coupling is expressed as \cite{Fischer:2003rp}:
\begin{eqnarray}
\alpha_{\rm fit}(p^2) &=& \frac{\alpha_S(0)}{1+p^2/\Lambda^2_{\tt QCD}}
+\frac{4 \pi}{\beta_0} \frac{p^2}{\Lambda^2_{\tt QCD}+p^2}
\left(\frac{1}{\ln(p^2/\Lambda^2_{\tt QCD})}
- \frac{1}{p^2/\Lambda_{\tt QCD}^2 -1}\right) 
\end{eqnarray}
with $\beta_0=(11N_c-2N_f)/3$.
It is important to note that the gluon propagator (\ref{fitII})
possesses a form such
that {\em Wick rotation is possible!}

\section{Dynamically induced scalar quark confinement}

\begin{figure}[th] 
\centerline{\psfig{file=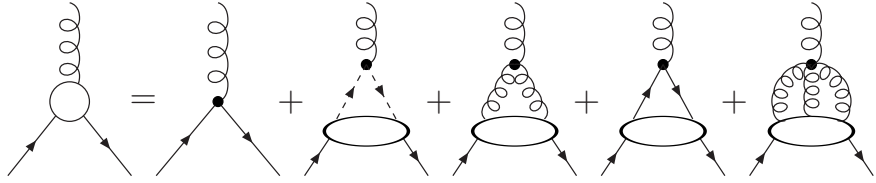,width=120mm}}
\vspace*{-8pt}
\caption{The
 Dyson-Schwinger equation  for the quark-gluon vertex.
\label{QGV}}
\end{figure}

\begin{figure}[th]
\centerline{\psfig{file=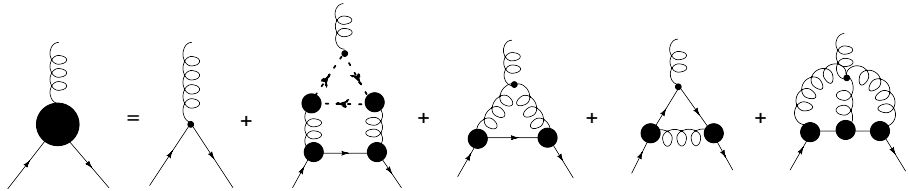,width=120mm}}
\vspace*{-8pt}
\caption{Some leading terms in the skeleton expansion for the quark-gluon vertex.
\label{QGV-skel}}
\end{figure}

The above presented results provide an explanation how gluon confinement  works in
a covariant gauge, but due to the infrared suppression of the gluon
propagator quark confinement seems even  more mysterious than ever.  To
proceed as in the above described studies the Dyson-Schwinger equation
for the quark propagator is analyzed with the result 
that the structure of the  quark
propagator depends crucially on the quark-gluon vertex \cite{Roberts:1994dr,Fischer:2003rp,Alkofer:2000wg,Fischer:2006ub,
Alkofer:2006jf}.
Therefore a detailed study of this three-point function, and especially its
infrared behaviour, is mandatory. Its Dyson-Schwinger equation is
diagrammatically depicted in fig.\ \ref{QGV}, its skeleton expansion in
fig.\ \ref{QGV-skel}.
But there is a drastic difference of the quarks as compared
to Yang-Mills fields: They possess a current mass.
{Even if this were not the case one expects dynamical chiral
symmetry breaking and thus dynamical mass generation to occur.} 

To  generalize the
infrared analysis of the Yang-Mills theory  to full QCD
\cite{Alkofer:2006gz} one concentrates first on the quark sector of
quenched QCD and chooses the masses of the valence quarks to  be large, {\it
i.e.\/}  $m > \Lambda_{\tt QCD}$.   The remaining scales below $\Lambda_{\tt
QCD}$ are those of the external momenta of the propagators and vertex functions. 
%The relevant infrared limit is the one where all these external momenta approach
%zero. 
Then the Dyson-Schwinger equations can be used
to determine the selfconsistent solutions  in
terms of powers of the small external momentum scale $p^2 \ll \Lambda_{\tt
QCD}^2$. The equations which have to be considered in addition to the ones of
Yang-Mills theory are the one for the quark propagator and the quark-gluon
vertex.

The full quark-gluon vertex $\Gamma_\mu$ can consist of up to twelve
linearly independent Dirac tensors. Some of those would vanish if chiral
symmetry would be realized in the Wigner-Weyl mode: These
tensor structures can be non-vanishing either if chiral symmetry is explicitely
broken by current masses and/or chiral symmetry is realized in the Nambu-Goldstone
mode ({\it i.e.} spontaneously broken). From a solution of the Dyson-Schwinger
equations we infer that  these ``Dirac-scalar'' structures are, in the chiral
limit, generated non-perturbatively together with the dynamical quark mass
function in a self-consistent fashion: Dynamical chiral symmetry breaking
reflects itself not only in the propagator but also in the quark-gluon vertex
function.

From such an infrared analysis one obtains
an infrared divergent solution for the quark-gluon vertex
such that  Dirac vector and {\em ``scalar''}
components of this  vertex are infrared divergent with exponent $-\kappa -
\frac 1 2$ \cite{Alkofer:2006gz}. A numerical solution of a truncated set of
Dyson-Schwinger equations confirms this infrared behavior. The
driving pieces of this solution are the scalar Dirac amplitudes of the
quark-gluon vertex and the scalar part of the quark propagator. Both pieces are
only present when chiral symmetry is broken, either explicitely or dynamically.

For the coupling related to the quark-gluon vertex
we obtain
\begin{equation}
\alpha^{qg}(p^2) = \alpha_\mu \,
{ [\Gamma^{qg}(p^2)]^2} \, { [Z_f(p^2)]^2}\,
{ Z(p^2)} \sim  \frac{const_{qg}}{N_c} \frac{1}{p^2} ,
\label{aq}
\end{equation}
 using that 
 \begin{equation}
{\Gamma^{qg}(p^2) \sim (p^2)^{-1/2-\kappa}} \,, \ \
{ Z_f(p^2) \sim const}\,, \ \ { Z(p^2) \sim (p^2)^{2\kappa}} .
\end{equation}
Note that  the coupling (\ref{aq}) is  singular in the infrared contrary to the
couplings  from the Yang-Mills vertices.

In a next step the anomalous
infrared exponent of the four-quark function is determined.  
Note that the static quark potential can be obtained
from this four-quark one-particle irreducible Greens function, which, including
the canonical dimensions, behaves like $(p^2)^{-2}$ for $p^2\to0$.
Therefore employing the well-known relation for a function $F\propto (p^2)^{-2}$
one obtains
\begin{equation}
V({\bf r}) = \int \frac{d^3p}{(2\pi)^3}  F(p^0=0,{\bf p})  e^{i {\bf p r}}
\ \ \sim \ \ |{\bf r} |
\end{equation}
for the static quark-antiquark potential $V({\bf r})$.
We conclude at this point that, given the infrared divergence of the
quark-gluon vertex as found in the solution of the coupled system of
Dyson-Schwinger equations, the vertex overcompensates the infared suppression
of the gluon propagator, and one therefore obtains a linear rising potential.
In addition, this potential is dynamically induced and has a strong scalar component.

To elucidate the here found relation between chiral symmetry breaking and 
quark confinement we keep chiral symmetry 
artificially in the Wigner-Weyl mode, {\it i.e.} in the chiral limit we force
the quark mass term as well as the ``scalar'' terms in the quark-gluon vertex
to be zero. We then find that 
the  resulting running coupling from the quark-gluon vertex is no longer
diverging but goes to a fixed point in the infrared similar to the couplings
from the Yang-Mills vertices. Correspondingly, one obtains 
a $1/r$ behaviour of the static quark potential.
The ``forced'' restoration of chiral symmetry is therefore directly linked with
the disappearance of  quark confinement. The infared properties of the
quark-gluon vertex in the ``unforced'' solution thus constitute a novel
mechanism that  directly links chiral symmetry breaking with confinement.

\section{Summary}

In this talk we have reported on results of functional approaches to
infrared QCD in the Landau gauge. We have elucidated
the mechanism for gluon confinement: Positivity of transverse gluons is
violated. Furthermore,
in the Yang-Mills sector the strong running coupling is infrared finite
whereas the running coupling from the quark-gluon vertex is infrared divergent.
Chiral symmetry is dynamically broken, and this takes place
in the quark propagator and the quark-gluon vertex.
  We have provided clear evidence that static quark confinement 
  in the Landau gauge is
  due to the infrared divergence of the quark-gluon vertex. 
  In the infrared this vertex has strong scalar components which induce a
  relation between confinement and broken chiral symmetry.

\section*{Acknowledgements}

RA thanks the organisers of {\it Lattice 2007\/}
for all their efforts which made this extraordinary conference possible.
We are grateful to A.~Cucchieri, A.~Maas, T.~Mendez,  J.~Pawlowski,
J.~Skullerud, and L.~v.~Smekal 
for interesting discussions. 
This work was supported by the DFG under grant no.\ Al 279/5-2, 
by the Helmholtz-University Young Investigator Grant VH-NG-332, 
by the FWF  under contract M979-N16, and by MEC travel grant PR2007-0110, Spain.

\end{document}